\documentclass[12pt]{article}
\begin{document}
\thispagestyle{empty}
\begin{center}

{\Large\bf Dispersion relations and  Omn\`es representations for
$K\to \pi\pi$ decay amplitudes}
\vskip1.4cm
C. Bourrely $^a$,  I. Caprini $^b$ and L. Micu  $^b$
\vskip0.3cm
$^a$ Centre de Physique Th\'eorique\footnote{Unit\'e propre de  Recherche 
7061}, CNRS-Luminy, \\Case 907, F-13288 Marseille Cedex 9 - France
\vskip 0.2cm
$^b$ National Institute of Physics and Nuclear Engineering,\\
POB MG 6, Bucharest, R-76900 Romania
\vskip 2cm
{\bf Abstract}\end{center}
We derive  dispersion relations for  $K\to\pi\pi$ decay,
using the Lehmann-Symanzik-Zimmermann formalism, which allows the
analytic continuation of the amplitudes with respect to the  momenta of the 
external particles. No off-shell extrapolation of the field operators is 
assumed. We obtain generalized Omn\`es representations, which incorporate 
the $\pi\pi$ and $\pi K$ $S$-wave phase shifts in the elastic region of the 
direct and crossed channels, according to Watson theorem.
The contribution  of the inelastic final-state and initial-state interactions
is parametrized by the technique of  conformal mappings.
We compare our results with previous dispersive treatments
and indicate how the formalism can be combined  with lattice calculations 
to yield  physical predictions.
\vskip 3cm
\noindent CPT-2002/P.4432
\newpage
\setcounter{page}{1}
\section{Introduction}

The  weak decay  $K\to \pi\pi$ has been a continuous challenge for the
theoretical investigations. Chiral perturbation theory (ChPT) was extensively 
used \cite{ChPT}, but the large number of unknown counterterms and 
renormalization constants render the numerical predictions difficult beyond 
the leading order.
Most lattice calculations (see \cite{Sach} and references therein)  simulate 
matrix elements of the type $\langle \pi | O_i| K\rangle$, related to 
the physical matrix elements $\langle \pi\pi | O_i| K\rangle$ by lowest 
order ChPT \cite{Bern}.
In this procedure the higher order final state interactions  are completely 
missing, while it is expected that they  play an important role for the 
$\Delta I=1/2$ rule and the CP-violating ratio $\epsilon'/\epsilon$.  
The finite-volume techniques developed in \cite{LeLu} can take into account
FSI, but they are numerically very demanding \cite{Sach}. In a combined 
approach  proposed recently, the results obtained by lattice simulations
at unphysical points are extrapolated to the physical configuration
by using calculations to NLO in ChPT \cite{Bouc, Lin}.

An alternative way to connect the on-shell amplitude to lattice results at
unphysical points and to spectral functions measured experimentally is based 
on dispersion relations. This formalism was  used some time ago for the
 CP-conserving amplitudes in order to explain the $\Delta I=1/2$ rule 
\cite{Truong}, and more recently in \cite{PaPi, PaPi1} for evaluating
the effects of final state  interactions  upon  $\epsilon'/\epsilon$. 
The last works use an Omn\`es representation \cite{Omnes} for the decay
amplitude, written by analogy with the case of the scalar form factor.
This approach was investigated further in Refs. \cite{Buras}-\cite{Cola},  
where  some critical remarks about the method were advanced.
An alternative dispersive framework for  $K\to \pi\pi$ decay
was proposed in \cite{BuCo1}, by assuming that the weak hamiltonian
carries a non-zero momentum. Then the matrix element of the decay
becomes equivalent with the $\pi K$ elastic scattering amplitude, for which 
Mandelstam representation is assumed.

In the references mentioned above, the dispersion relations for the weak 
decay were written by  using the analogy with the familiar cases of the form 
factors or the scattering amplitudes. However, in the weak decay
a continuation in the external momenta is necessary in order to obtain a 
dispersion relation. As a proof of the dispersion relations in this case
is missing, their meaning was not always clear and led to some confusion. 
The dispersive variable was interpreted either as the mass
or the momentum of an off-shell particle. The clarification of this point is 
possible only by a systematic derivation in the frame of a field theoretic 
formalism.
In the present work we address this problem, by performing the continuation 
in the external momenta with the Lehmann-Symanzik-Zimmermann (LSZ) formalism 
\cite{LSZ}.  
Our main result is  a general Omn\`es representation for the 
$K\to \pi\pi$ amplitudes, including final and initial state  interactions 
in both the direct ($K\to\pi\pi$) and the crossed ($\pi\to\pi K$) channels. 
The derivation clarifies the significance of the dispersive variables, 
allowing to make contact with lattice calculations done at unphysical points.

In the next section we present the derivation of the dispersion relations,
using LSZ reduction and hadronic unitarity. We follow  to some extent
the dispersive treatment of $B\to \pi\pi$ decay considered in 
\cite{CaMi, CaMi1}. However, the different masses of the decaying particles in
the two processes require specific treatments. In section 3, we derive a
generalized Omn\`es representation for $K\to \pi\pi$ decay  by solving 
the inhomogeneous Hilbert problem \cite{Musk, PhTr} in the direct and 
the crossed channels.
In Section 4, we compare our results with the dispersion relations
considered previously in the literature, and show how to combine them with 
lattice calculations in order to predict the physical amplitude.
\section{LSZ reduction and dispersion relations}

We consider decay amplitudes $A_I$ of definite isospin, $I=0, 2$, defined as
\begin{equation}\label{A}
A_I =  \langle (\pi(k_1)\,
\pi(k_2))_I\,;{\rm out\,}|{\cal H}_w(0)|K(p)\,; {\rm in}\rangle\,,
\end{equation}
where the ``in'' and ``out\,'' states  are defined with respect to the strong
interactions and ${\cal H}_w$ is the weak effective hamiltonian density
\cite{Hweak}
\begin{eqnarray}\label{Hw}
&&{\cal H}_w(x)={G_F\over \sqrt{2}}\sum\limits_{k=u,c} V_{kd} V^*_{ks}  \\
&&\times\left[
C_1(\mu) O_1^k(x,\mu)+C_2(\mu)O_2^k (x,\mu) +\sum\limits_{j=3,\ldots ,8}
C_j(\mu) O_j(x, \mu)\right]\,.\nonumber
\end{eqnarray}
Here $O_j$ are local $\Delta S=1$, $\Delta B=0$ operators and
$C_j$ the corresponding Wilson coefficients, which take into account
perturbatively the strong dynamics at distances shorter than $1/\mu$. We assume
that a factor $i$ was included in the definition of the operators, so that 
the amplitudes (\ref{A}) satisfy time reversal invariance, up to the 
complex coefficients in (\ref{Hw}).

For our purpose it is more convenient to start from the $S$-matrix element 
\begin{equation}\label{defS}
S_I=\langle \pi(k_1)\, \pi(k_2); {\rm out\,}| K(p); {\rm in}\rangle\,,
\end{equation}
where the transition from the ``in'' to the ``out'' states is achieved by
both the strong and weak interactions.
By expanding the $S-$matrix to first order in the weak interactions
one obtains the expression (\ref{A}) of the decay amplitude.
Alternatively, by applying the LSZ reduction \cite{LSZ} to the $K$
meson in Eq. (\ref{defS}), the decay amplitude (\ref{A}) is expressed as
\begin{equation}\label{AI1}
A_I = {1\over \sqrt{2
p_0}}\langle \pi(k_1)\, \pi(k_2);  {\rm out\,}|
\,\eta_{K}(0)\,|0\rangle \,,
\end{equation}
where $\eta_{K}(x)={\cal K}_x \phi_{K}(x)$ denotes the source  
operator (${\cal K}_x $ is the Klein-Gordon operator and $\phi_{K}$ the
interpolating field of the kaon).
In a Lagrangian theory the source operator has the formal expression
\begin{equation}\label{etaK}
\eta_{K}(x)={\delta {\cal L}_{int}\over \delta
\phi_{K}}-\partial_\mu{\delta
{\cal L}_{int}\over \delta \partial_\mu\phi_{K}}\,,
\end{equation} 
{\it i.e.}  it has contributions from both the strong and weak parts of
the interaction Lagrangian. In what follows we do not need the explicit 
expressions of the sources, but only the significance of the matrix elements 
involving them. We stress also that throughout the derivation the sources 
are on-shell operators, defined in terms of the physical interpolation fields.

The matrix element (\ref{AI1}) depends on the momenta $k_1$ and $k_2$ of the 
two pions.
We  shall consider it as a function of the invariant variables 
$s=(k_1+k_2)^2$, $t=k_1^2$, $u=k_2^2$.
The physical amplitude  corresponds to  the values
$s=m_K^2$, $t=m_\pi^2$ and $u=m_\pi^2$.
The extrapolation to arbitrary external momenta can be achieved by the LSZ
reduction formalism \cite{LSZ}. We remark that Eq. (\ref{AI1}) is similar to
the definition of the electromagnetic form factor of the pion, where $\eta_K$
is replaced by the electromagnetic current $J_\mu$.
We can apply therefore the standard methods used in deriving
the dispersion relations for the pion form factor \cite{Bart}.
Making the LSZ reduction of one final pion in Eq. (\ref{AI1}), we obtain
\begin{equation}\label{lsza}
A_I(s,t) ={i\over \sqrt{4
k_{10} p_0}}\int {\rm d}x e^{ik_1 x} \theta (x_0) \,\langle
\pi(k_2)|[\eta_{\pi}(x), \eta_{K} (0)]|0\rangle \,,
\end{equation}
where $\eta_{\pi}(x)$ is the source of the reduced  pion.
We left aside the so-called "degenerate terms'' which  are polynomial of
the Lorentz invariant variables \cite{Bart}. Then Eq. (\ref{lsza}) defines
a function holomorphic for those values of the external squared momenta
$s$ and $t$ for which the integral is convergent (for the unreduced pion we
take the physical value $u=m_\pi^2$). Due to the presence of $\theta(x_0)$,
the integral upon $x_0$ converges  in the upper half of the $k_{10}$ complex
plane, ${\rm Im}\, k_{10}>0$. 
The causality property of the commutator restricts the integral 
upon the spatial variables to $|{\bf x}|< |x_0|$ \cite{Bart}.
We choose the particular Lorentz frame where the unreduced pion is at rest
(${\bf k_2}=0$), when  $k_{10}=(s-t-m_\pi^2)/ 2 m_\pi$ and
${\bf k}_1^2 =[s-(\sqrt{t}+m_\pi)^2][s-(\sqrt{t}-m_\pi)^2]/(2 m_\pi)^2$.
Then the integral in (\ref{lsza}) represents a function of $s$ and $t$, 
analytic for complex values of these variables, with possible
discontinuities along the real axes. The rigorous proof  of the analyticity
in the external masses is actually a difficult problem and requires a more 
detailed analysis \cite{KaWi}.
Here we do not attempt to give a proof, but only use the LSZ representation
to understand the meaning of the dispersive variables and to read off the 
contributions to the spectral functions appearing in the dispersion relation.

The discontinuity across the real axis is obtained formally from the
expression (\ref{lsza}) by replacing $i\theta(x_0)$ by $1/2$, inserting a 
complete set of intermediate states in the commutator
$[\eta_{\pi}(x), \eta_{K} (0)]$ and using translational invariance \cite{Bart}.
The two terms in the commutator allow us to decompose the spectral function as
\begin{equation}\label{sigma}
\sigma = \sigma_s   + \sigma_t \,,
\end{equation}
where
\begin{equation}\label{sigmas} \sigma_s={1\over 2 \sqrt{4 k_{01}p_0}}
\sum_{n}\delta  (k_1+k_2-p_n) \langle \pi(k_2)|
\eta_{\pi}(0)|n\rangle\langle n| \eta_{K}(0)|0\rangle\,,
\end{equation} and
\begin{equation}\label{sigmat}
\sigma_t = {1\over 2
\sqrt{4 k_{01}p_0}}\sum_{n}\delta  (k_1+p_n)  \langle \pi(k_2)|
\eta_{K}(0)|n\rangle\langle n| \eta_{\pi}(0)|0\rangle\,.
\end{equation}
In these relations the summation is over intermediate states consisting of 
physical particles, with an implicit integration upon their momenta.
By the subscripts $s$ ($t$), we anticipate the fact that $\sigma_s$ receives 
contributions from the $s-$channel ($K\to \pi\pi$) and $\sigma_t$ from the
$t-$channel ($\pi\to K\pi$).
Accordingly, we can write the amplitude as a sum of two terms,
$A_s^I$ and $A_t^I$, obtained by a dispersion representation
involving the spectral function $\sigma_s$  and $\sigma_t$, respectively.
In order to evaluate the spectral functions, we recall that 
the sources contain contributions from both the strong and the weak
interactions, the last ones being treated to first order.

Let us consider first the  spectral function $\sigma_s$ defined in  
(\ref{sigmas}). As discussed in Ref. \cite{CaMi1}, the intermediate states $n$
which contribute to the unitarity sum are generated by either the weak
or the strong part of the  source  $\eta_K$, undergoing a rescattering 
to the final $\pi\pi$ state by a strong (weak) process, respectively.
The first contributions represent the so-called  "final state
interactions" (FSI), while the second  are usually interpreted as
"initial state interactions" (ISI).
The lowest intermediate state contributing to FSI consists of
two-pions, which produces the  branch-point $s=4m_\pi^2$, while
for the ISI the lowest intermediate state is the pair $K^* \pi$, responsible 
for the threshold $s=(m_{K^*}+m_\pi)^2$. 
In order to write the specific contributions, we recall that
in the LSZ formalism  the matrix elements of the sources
represent, up to kinematical factors, the physical decay or scattering
amplitudes \cite{Bart}. Thus, according to  Eq. (\ref{AI1}),
$\langle \pi\pi|\eta_{K}(0)|0\rangle$ is the amplitude of the weak
decay of $K$ into two pions (here only the weak part of the source
contributes), while $\langle \pi(k_2)| \eta_{\pi}(0)|n\rangle$
is the amplitude of either the strong  or the weak  $n\to \pi\pi$ transition,
depending on which part of the source is considered.

A remarkable property of $\sigma_s$ is that it depends only on $s$, being
independent of the variable $t$ \cite{CaMi1}, \cite{Bart}. This can be easily
seen by recalling that the intermediate states $|n\rangle$ consist of
physical particles. By choosing the c.m. system, where $p_n^2=s$ is the total
energy squared, we see that the matrix elements in (\ref{sigmas})
depend only on $s$ and the physical masses. 
In the two-particle approximation of the unitarity sum, the integral in
(\ref{sigmas}) can be performed exactly \cite{CaMi}.
According to the discussion above we can write
\begin{equation}\label{sigmas1}
\sigma_s(s) = \sigma_{FSI}(s)   + \sigma_{ISI}(s) \,,
\end{equation}
where the first contributions to each term are
\begin{eqnarray}\label{sfsi}
\sigma_{FSI}(s)&=&\theta(s- 4m_\pi^2) M^*_{\pi\pi\to \pi\pi} A_{K\to\pi\pi}+
\theta(s-4m_K^2) M^*_{\bar K K\to \pi\pi}  A_{K\to\bar K K}+\ldots\,,
\nonumber\\
\sigma_{ISI}(s)&=&\theta(s- (m_{K^*}+m_\pi)^2)
A^*_{K^*\pi\to \pi\pi}  M_{K\to K^* \pi}(s) +\ldots \,.
\end{eqnarray}
In the above relations $M_{\pi\pi\to \pi\pi}$ and $M_{\bar K K\to \pi\pi}$
denote on-shell $S$-wave strong scattering amplitudes at c.m. energy squared 
$s$, and $A_{K\to \pi\pi}$ ($ M_{K\to K^* \pi}$), etc., are weak (strong)  
decay amplitudes.
The amplitude $A_s^I$ can be recovered from the discontinuity by means of a
dispersion integral. Neglecting for the moment possible subtractions and
polynomials in the Mandelstam variables, we have
\begin{equation}\label{drs}
A_s^I(s) =
{1\over \pi }\int\limits_{4 m_\pi^2}^\infty {\sigma_{FSI} (s')
\over s'-s} {\rm d}s' +{1\over \pi } \int\limits_{(m_{K^*}+m_\pi)^2}^\infty
{\sigma_{ISI}(s') \over s'-s} {\rm d}s'\,.
\end{equation}

We consider now the  spectral function $\sigma_t$ defined in (\ref{sigmat}).
The weak part of the source $\eta_\pi$ is responsible for the FSI in the 
$t$-channel, with the lowest branch-point at $t=(m_K+m_\pi)^2$.
The strong part of the source $\eta_\pi$ generates the ISI in the $t$-channel, 
with the lowest branch-point $t=9 m_\pi^2$.
As above, it is easy to show that $\sigma_t$ does not depend on $s$ and
can be written as
\begin{equation}\label{sigmat1}
\sigma_t(t) = \tilde\sigma_{FSI}(t)+\tilde \sigma_{ISI}(t) \,,
\end{equation}
where, according to the above discussion, the lowest terms are
\begin{eqnarray}\label{tfsi}
\tilde\sigma_{FSI}(t)&=&\theta(t-(m_K+m_\pi)^2)
N^*_{\pi K\to \pi K} A_{\pi \to \pi K}+\ldots\,,\nonumber\\
\tilde\sigma_{ISI}(t)&=& \theta(t- 9 m_\pi^2)\, N^*_{\pi \to 3 \pi}
 A_{ 3 \pi\to \pi K} +\ldots\,.
\end{eqnarray}
Here $ N_{\pi K\to \pi K}$ is the $\pi K$  $S$-wave 
scattering amplitude at c.m. energy squared equal to $t$, 
and $A_{\pi \to \pi K}\,(N_{\pi \to 3 \pi})$ etc., are weak (strong) decay 
amplitudes.
Neglecting again possible subtractions, we write the amplitude $A_t^I$
in terms of its discontinuity in the $t$-channel by a dispersion integral
\begin{equation}\label{drt}
A_t^I(t)= {1\over \pi}
\int\limits^\infty_{(m_K+ m_\pi)^2} {\tilde\sigma_{FSI} (t') \over t'
-t} {\rm d} t'\,+{1\over \pi}
\int\limits^\infty_{9 m_\pi^2} {\tilde\sigma_{ISI} (t') \over t'
-t }{\rm d} t'\,\,.
\end{equation}
The total amplitude $A_I(s,t)$ is then expressed as the sum
\begin{equation}\label{Ast}
A_I(s,t)=A_s^I(s)+A_t^I(t)\,,
\end{equation}
the physical amplitude being obtained for $s=m_K^2$ and $t=m_\pi^2$.
We note in particular that $A_t^I(m_\pi^2)$ is a real number, since the
point $t=m_\pi^2$ is situated below the cuts in the dispersion relation 
(\ref{drt}).

The significance of the variables $s$ and $t$ is clear from the
above discussion: $s$ is defined in terms of the
pion momenta as $s=(k_1+k_2)^2$ and $t$ is equal to the external
momentum squared of one pion, $t=k_1^2$. Therefore, it represents
the mass squared of one external pion. We recall that in the above
formalism no off-shell extrapolation was assumed, the sources entering
the matrix elements being on-shell operators.

\section{Omn\`es representations}

It is convenient to write the above
dispersion relations in terms of the phases of the rescattering
amplitudes in the elastic region, according to Watson theorem \cite{Watson}.
To illustrate the method, we consider first the amplitude $A_I(s, m_\pi^2)$ 
as a function of $s$ at physical $t=m_\pi^2$. 
The general case of $A_I(s,t)$ will be treated in subsection \ref{sec32}.
This generalization is useful in order to incorporate information available 
on the decay amplitude at nonphysical pion masses.
\subsection{Amplitude  $A_I(s, m_\pi^2)$} 
\label{sec31}

In this case, as mentioned above, the last term in  Eq. (\ref{Ast}) is a real
constant. Denoting $A_\pm = A_I(s\pm i \epsilon, m_\pi^2)$,
we write the unitarity relation in the $s$-channel as
\begin{equation}\label{units}
{A_+-A_-\over 2 i}= \theta (s-4 m_{\pi}^2) M ^*_I(s) A_++
\theta (s-s_{in})\,\sigma_{in}(s)\,,
\end{equation}
where 
\begin{equation}\label{M0}
M_I (s)={\eta_0^I e^{2 i \delta_0^I}-1\over 2 i}\,,
\end{equation}
is the S-wave $\pi \pi$ scattering amplitude of isospin $I$. In Eq. 
(\ref{units}) $\sigma_{in}$ denotes the sum of the inelastic FSI and the ISI
spectral functions (we take $s_{in}$ equal to the FSI inelastic branch-point 
$4 m_K^2$, which is lower than the ISI branch-point 
$(m_{K^*}+m_\pi)^2$). In the r.h.s. of Eq. (\ref{units}) we note the presence 
of the amplitude $A_+=A_I (s+i\epsilon, m_\pi^2)$, due to the fact that 
the intermediate two pions in the unitarity sum are physical particles, 
as we mentioned above.

The relation (\ref{units}) can be written as an inhomogeneous Hilbert
equation \cite{Musk}
\begin{equation}\label{H} A_{+}(1- 2i M^*_I)- A_{-}= 2 i\, \theta (s-s_{in})\,
\sigma _{in}(s)\,, \quad s\ge 4 m_\pi^2\,.
\end{equation}
We shall construct the solution by imposing time reversal invariance, which
implies that the amplitudes satisfy the reality condition $A_I(s^*)=A^*_I(s)$ 
and the discontinuity across the cut is equal to the imaginary 
part \cite{PhTr}. Using the expression (\ref{M0}) we obtain from (\ref{H}), 
for $s\ge 4 m_\pi^2$,
\begin{equation}
\label{H1}
{\rm Im} A_I \cos \delta_0^I- {\rm Re}  A_I \sin \delta_0^I=
\theta (s-s_{in}){2\, {\rm Re} \, 
[\sigma_{in} e^{i \delta_0^I}]\over 1+\eta_0^I}\,.
\end{equation}
We define now the Omn\`es function
\begin{equation}\label{Omegas}
\Omega_{I}(s)=\exp\left[{s-s_0\over \pi}
\int\limits_{4 m_\pi^2}^\infty {\delta_0^I(s') \,{\rm d}s'\over
(s'-s_0)(s'-s)}\right]\,,
\end{equation}
assuming that one subtraction is sufficient. The boundary values of 
$\Omega_I(s)$ satisfy the relations $ \Omega_I(s\pm i\epsilon) =
\exp(\pm i \delta_0^I) |\Omega_I(s)| $ (we recall that the modulus 
$|\Omega_I(s)|$ is obtained from (\ref{Omegas}) by taking the 
principal value of the integral).
Then Eq. (\ref{H1}) can be written as
\begin{equation}
\label{ratio} {\rm Im}\left [{A_I(s, m_\pi^2) \over \Omega_{I}(s)}\right]
= \theta (s-s_{in})  {2 \over 1+\eta_0^I}\,{  {\rm Re}\,[\sigma_{in}\, 
e^{i \delta_0^I}]\over |\Omega_I (s)|}\,.
\end{equation}
We define now the function $G_I(s)$ through the relation
\begin{equation}
\label{AOmG} 
A_I(s, m_\pi^2)=\Omega_I (s)\,G_I (s)\,,
\end{equation}
and express it by a dispersion relation in terms of its imaginary part given in
(\ref{ratio}):
\begin{equation}
\label{Gdr}
G_I(s) =  P_I(s)+ {s-s_0\over \pi}  \int\limits_{s_{in}}^\infty {\rm d} s'
{2\over 1+\eta_0^I}\,{{\rm Re}\,[ \sigma _{in}(s')e^{i\delta_0^I(s')}]
\over |\Omega_I(s')|\,  (s'-s_0)(s'-s) }\,.
\end{equation}
Here $P_I(s)$ is a polynomial and, for convenience,  we wrote the integral 
with one subtraction. The subtractions are actually not relevant in our method,
since we shall parametrize the function $G_I(s)$ in a different way, using the
technique of conformal mappings\footnote{The method of conformal mappings was
proposed in particle physics a long time ago \cite{CiFi}.
In a  context similar to the present one, the method was applied to the pion 
electromagnetic form factor in Ref. \cite{HeLa}.}. 
Namely, since  by construction $G_I(s)$ is  analytic
in the $s-$plane cut for $s>s_{in}$, we consider the variable
\begin{equation}
\label{z}
z(s)={\sqrt{s_{in}-m_\pi^2}-\sqrt{s_{in}-s}\over \sqrt{s_{in}-m_\pi^2} +
\sqrt{s_{in}-s}}\,,
\end{equation}
which maps the $s$-plane cut along the real axis for $s>s_{in}$ onto the disk 
$z<1$ of the plane $z=z(s)$. Actually, the mapping of the $s$-plane onto the
unit disk is not unique \cite{HeLa}. For further convenience we choosed 
the mapping such that  $z(m_\pi^2)=0$. Now we expand $G_I(s)$
in powers the variable $z$
\begin{equation}
\label{Gz}
G_I(s)=\sum\limits_n a_n^{(I)}\, [z(s)]^n\,,
\end{equation}
where $a_n^{(I)}$ are real numbers. This series converges in the whole disk 
$|z|<1$, {\em i.e,} in the whole $s$-plane cut along $s>s_{in}$, in particular
at the  physical point $s=m_K^2$.

Inserting (\ref{Gz}) into (\ref{AOmG}) we obtain a representation of the
amplitude
\begin{equation}
\label{AGz}
A_I (s, m_\pi^2)= \Omega_{I}(s)\, \sum\limits_n a_n^{(I)} [z(s)]^n\,,
\end{equation}
in terms of the known $S-$wave $\pi\pi$ phase shifts entering the Omn\`es 
function $\Omega_{I}(s)$, and the real Taylor coefficients
$a_n^{(I)}$. In Section 5 we shall discuss  how these coefficients can be
determined by using lattice results at unphysical values of $s$. 
The physical amplitude is obtained from (\ref{AGz}) by setting $s=m_K^2$.
\subsection{Amplitude $A_I(s,t)$}
\label{sec32}

We consider now the amplitude $A_I(s, t)$ for arbitrary arguments.
It turns out that the elastic unitarity for
$A_I(s, t)$ can not be immediately solved by means of an Omn\`es
representation, as in the above treatment of $A_I (s, m_\pi^2)$. Indeed,
from the relations (\ref{Ast}) and (\ref{sigmas1}-\ref{drs}) the
discontinuity of $A_I(s, t)$ across the cut along $s>4 m_{\pi}^2$ at 
fixed $t$ is
\begin{equation}
\label{unitst}
{A_I (s+i\epsilon, t)-A_I (s-i\epsilon, t)\over 2 i}= M ^*_I(s) 
A_I(s+i\epsilon, m_\pi^2)+ \theta (s-s_{in})\sigma_{in}(s)\,.
\end{equation}
We note the presence, in the r.h.s., of the amplitude 
$A_I (s+i\epsilon, m_\pi^2)$, due to the fact that the intermediate states 
in the unitarity sum consist of physical particles.
Therefore the functions which appear on the two sides of (\ref{unitst}) are 
different, and this relation can not be written as a Hilbert boundary value 
equation.

This difficulty can be circumvented if we treat separately the functions
$A_s^I(s)$ and $A_t^I(t)$ defined in Eqs. (\ref{drs}) and (\ref{drt}), 
respectively.
Denoting $A_{s,\pm} = A_s^I(s\pm i \epsilon)$ we obtain from Eqs. 
(\ref{sigmas1}-\ref{drs})
\begin{equation}
\label{unitAs}
{A_{s,+}-A_{s,-}\over 2 i}= \theta (s-4 m_{\pi}^2) M ^*_I(s) A_I(s+ i
\epsilon, m_\pi^2)+ \theta (s-s_{in})\,\sigma_{in}(s)\,.
\end{equation}
In order to bring this relation to a form convenient for the
Muskhelishvili-Omn\`es technique, we express, according to Eq. (\ref{Ast}),
\begin{equation}
\label{replace}
A_I(s+ i \epsilon, m_\pi^2)= A_{s, +}+A_t^I( m_\pi^2)\,.
\end{equation}
Then Eq. (\ref{unitAs}) becomes:
\begin{equation}
\label{Hs} 
A_{s,+}(1- 2i M^*_I)- A_{s,-}= 2 i\,[\theta (s-
4 m_\pi^2) M^*_I A_t^I(m_\pi^2)+ \theta (s-s_{in})\, \sigma _{in}(s)]\,.
\end{equation}
This equation is similar to Eq. (\ref{H}), excepted for an additional term in 
the r.h.s., which contributes to the imaginary part above the elastic 
threshold $4 m_\pi^2$. Therefore, $A_s(s)$ will be of the form (\ref{AOmG}), 
with the function  $G_I(s)$ given by a dispersion
relation similar to (\ref{Gdr}), containing in addition the term
\begin{equation}
\label{additio}
A_t^I(m_\pi^2)\,{s-s_0\over \pi}
\int\limits_{4 m_\pi^2}^\infty  {\rm d} s' {2\over 1+\eta_0^I }{{\rm Re}\,
[M^*_I (s')e^{i \delta_0^I(s') }] \over  |\Omega_I(s')| (s'-s_0)(s'-s)}\,,
\end{equation}
where we took into account that $A_t^I(m_\pi^2)$ is a real constant, 
as mentioned below Eq. (\ref{Ast}).

It is convenient to separate  in this integral the contribution of 
the inelastic  region $s> s_{in}$,  combining it
with the contribution of the inelastic term $\sigma_{in}$ and
expanding them in powers of the variable $z$ defined in (\ref{z}).
Therefore, we express $A_s^I (s)$ as
\begin{equation}
\label{AsGz}
A_s^I (s)= \Omega_{I}(s) \left[A_t^I(m_\pi^2)\, f_I(s)  
+\sum\limits_n c_n^{(I)}\, [z(s)]^n\right]\,,
\end{equation}
with
\begin{equation}
\label{f}
f_I(s)={s-s_0\over \pi}
\int\limits_{4 m_\pi^2}^{s_{in}} {\sin \delta_0^I (s') \,{\rm d} s'\,
\over  |\Omega_I(s')| (s'-s_0)(s'-s) }\,,
\end{equation}
where we took along the elastic region $\eta_0^I(s)=1$ and 
$M_I= e^{i \delta_0^I}\, \sin \delta_0^I $.

It is convenient to choose the subtraction point $s_0=m_\pi^2$ in both the
expression (\ref{Omegas}) of the  Omn\`es function and the definition (\ref{f})
of the function $f_I(s)$. This implies $\Omega_I(m_\pi^2)=1$ and
$f_I(m_\pi^2)=0$.
By recalling also that the conformal mapping (\ref{z}) was defined such as 
$z(m_\pi^2)=0$,   it follows that the amplitude $A_s^I(s)$ given by
Eq. (\ref{AsGz}) is normalized as
\begin{equation}
\label{Asmpi}
A_s^I(m_\pi^2)= c_0^{(I)}\,.
\end{equation}

We consider now the second term in the decomposition (\ref{Ast}), namely the
amplitude $A_t^I(t)$. In order to obtain an Omn\`es representation,
we must work with amplitudes  $\tilde A_J(s, t)$ of definite isospin in the 
$t$ channel, $\pi\to\pi K$. By crossing symmetry  we can write
\begin{equation}
\label{crossing}
A_I(s,t) =\sum\limits_{J={1\over 2},{3\over 2}} C_{I J} \tilde A_J(s,t)\,,
\quad I=0, 2\,,
\end{equation}
where the matrix $C_{IJ}$ is known from the elastic $\pi K$ scattering 
\cite{AnBu}.
Each amplitude  $\tilde A_J(s,t)$ admits a decomposition similar to (\ref{Ast})
\begin{equation}
\label{tildeAst}
\tilde A_J(s,t)=\tilde A_s^J(s)+\tilde A_t^J(t)\,.
\end{equation}
We consider the amplitude  $\tilde A_t^J(t)$ and define
$\tilde A_{t,\pm}= \tilde A_t^J (t\pm i\epsilon)$. Then the unitarity relation
in the $t$-channel, given by Eqs. ({\ref{sigmat1}-\ref{drt}), can be written as
\begin{equation}
\label{unitt}
{\tilde A_{t,+}-\tilde A_{t,-}\over 2 i}= \theta (t- (m_K+m_{\pi})^2)
N^*_J(t) \tilde A_J(m_K^2, t+i\epsilon)+
\theta (t-t_{in})\,\tilde\sigma_{in}(t)\,,
\end{equation}
where
\begin{equation}
\label{N0}
N_J(t)= {\tilde\eta_0^J e^{2 i \tilde\delta_0^J}-1 \over 2 i}\,,
\end{equation}
denotes the $S$-wave $\pi K$ scattering amplitudes at c.m. energy squared 
equal to $t$, and $\tilde\sigma_{in}(t)$ the contribution of the inelastic 
FSI and ISI $t$-channels. 
We take $t_{in}$ equal to the ISI branch-point $9 m_\pi^2$, which is lower 
than the inelastic FSI branch-point $(m_K+m_\eta)^2$.
 
Since the intermediate $\pi K$ state  in the unitarity sum (\ref{sigmat})
consists of physical particles, in the r.h.s. of (\ref{unitt})
contributes the amplitude $\tilde A_J(m_K^2, t+i\epsilon)$.
By expressing  $\tilde A_J(m_K^2, t+i\epsilon)$ according to (\ref{tildeAst})
we obtain
\begin{eqnarray}
\label{Ht}
\tilde A_{t,+}(1- 2i N^*_J)- \tilde A_{t,-}&=& 2 i\,[\theta (t-
( m_\pi+m_K)^2) N^*_J \tilde A_s^J(m_K^2)\nonumber\\
&+& \theta (t-t_{in})\, \tilde\sigma _{in}(t)]\,.
\end{eqnarray}
The solution of this equation can be obtained following the procedure applied 
to the function $A_s^I(s)$. We introduce the Omn\`es function
\begin{equation}
\label{Omegat}
\tilde\Omega_J(t)=
\exp\left[{t-t_0\over \pi} \int\limits_{(m_\pi+m_K)^2}^\infty
{ \tilde \delta_0^J(t')\over (t'-t_0)(t'-t)} {\rm d}t'\right]\,,
\end{equation}
and express the  ratio $\tilde A_t^J(t)/\tilde\Omega_J(t)$ through a 
dispersion relation in terms of its imaginary part calculated from (\ref{Ht}).
Then $\tilde A_t^J (t)$ can be written as
\begin{eqnarray}
\label{At}
\tilde A_t^J (t)&=& \tilde \Omega_J(t) \left[\tilde P_J(t)
+\,{t-t_0\over \pi}
\int\limits_{(m_\pi+m_K)^2}^{\infty}{\rm d} t' {2\over 1+\tilde \eta_0^J} 
{{\rm Re}\, [\tilde A_s^J(m_K^2)N^*_J(t') e^{i \tilde\delta_0^J}]
\over |\tilde\Omega_J(t')| (t'-t_0)(t'-t) }\right.\nonumber\\
&+&{t-t_0\over \pi} \left.   \int\limits_{t_{in}}^{\infty}{\rm d} t' 
{2\over 1+\tilde \eta_0^J}
{{\rm Re}\,[\sigma_{in}(t') e^{i \tilde\delta_0^J}]
\over  |\tilde\Omega_J(t')| (t'-t_0)(t'-t)}\right]\,.
\end{eqnarray}
We recall that the quantities  $ \tilde A_s^J(m_K^2)$ entering this relation
are complex numbers.

We further separate in Eq. (\ref{At}) the contribution of the elastic part of 
the cut, and take into account the higher singularities by means of 
a conformal mapping. Namely, we define the variable
\begin{equation}
\label{w}
w(t)={\sqrt{t_{in}-m_\pi^2} -\sqrt{t_{in}-t}\over\sqrt{t_{in}-m^2_\pi}-
\sqrt{t_{in}-t}}\,,
\end{equation}
which maps the $t$-plane cut for $t>t_{in}$ onto the disk $|w|<1$, 
such that $w(m_\pi^2)=0$, and expand the polynomial and the inelastic part 
of Eq. (\ref{At}) in powers of this variable:
\begin{equation}
\label{Gtw}
\tilde P_J+{t-t_0\over \pi}
\int\limits_{t_{in}}^{\infty}{\rm d} t' {2\over 1+\tilde \eta_0^J}
{{\rm Re} \{[\tilde A_s^J(m_K^2)\, N^*_J(t')+ \sigma_{in}(t')]
e^{i \tilde\delta_0^J}\} \over  |\tilde\Omega_J(t')| (t'-t_0) (t'-t)}
=\sum \tilde c_n^{(J)}\, [w(t)]^n\,,
\end{equation}
the coefficients $\tilde c_n^{(J)}$ being real. Then Eq. (\ref{At}) becomes
\begin{equation}
\label{AtGw}
\tilde A_t^J (t)= \tilde \Omega_{J}(t)\, \left [
{\rm Re }\tilde A_s^J(m_K^2)\, g_J(t)+\sum\limits_n \tilde c_n^{(J)}\, 
[w(t)]^n \right]\,,
\end{equation}
where we defined
\begin{equation}
\label{g}
g_J(t)= {t-t_0\over \pi}
\int\limits_{(m_\pi+m_K)^2}^{t_{in}} {\sin \delta_0^J\, {\rm d} t'
\over | \tilde\Omega_J(t')| (t'-t_0)(t'-t) }\,.
\end{equation}
We took into account the fact that in the elastic region
$\tilde \eta_0^J=1$ and
$\tilde N^*_J e^{i \tilde\delta}=\sin \tilde\delta_0^J$.

It is convenient to choose the subtraction point $t_0=m_\pi^2$ in both the 
Omn\`es function (\ref{Omegat}) and the definition (\ref{g}) of $g_J(t)$. 
This means that $\tilde\Omega_J(m_\pi^2)=1$ and  $g_J(m_\pi^2)=0$. 
Recalling also that the conformal variable $w(t)$ was defined in (\ref{w}) 
such that $w(m_\pi^2)=0$, we obtain from (\ref{AtGw})
\begin{equation}
\label{tildeAtmpi}
\tilde A_t^J(m_\pi^2)= \tilde c^{(J)}_0\,.
\end{equation}

Collecting Eqs. (\ref{Ast},\,{\ref{AsGz},\,\ref{crossing}) and
(\ref{AtGw}), we express the amplitude $A_I(s,t)$ as
\begin{eqnarray}
\label{Afinal}
A_I(s,t)&=& \Omega_{I}(s) \left[ A_t^I(m_\pi^2)\, f_I(s)
+\sum\limits_n c_n^{(I)} [z(s)]^n\right]\,\nonumber\\
&+& \sum\limits_J  C_{IJ}\,\tilde \Omega_{J}(t) \left [
{\rm Re }\tilde A_s^J(m_K^2)\, g_J(t)+\sum\limits_n \tilde c_n^{(J)} 
[w(t)]^n \right]\,,
\end{eqnarray}
where the functions $f_I(s)$ and $g_J(t)$ are defined in Eqs. (\ref{f})
and (\ref{g}), respectively (we recall that $s_{in}= 4 m_K^2$ and 
$t_{in}=9 m_\pi^2$).
In Eq. (\ref{Afinal}) we must insert, according to 
(\ref{Ast},\ref{crossing},\ref{tildeAst}) and (\ref{tildeAtmpi}),
\begin{equation}
\label{Atmpi1}
A_t^I(m_\pi^2) =\sum_J C_{IJ} \,\tilde c_0^{(J)}\,.
\end{equation}
Also, using the crossing relation (\ref{crossing}), we express
the quantity ${\rm Re}\, \tilde A_s^J(m_K^2)$ entering (\ref{Afinal}) as
\begin{equation}
\label{crossingt}
{\rm Re}\, \tilde A_s^J(m_K^2) =\sum\limits_{L=0,2} C_{ JL}^{-1} \,{\rm Re}\,
A_s^L(m_K^2)\,,\quad J={1\over 2}\,, {3\over 2}\,,
\end{equation}
where ${\rm Re}\, A_s^L(m_K^2)$ is obtained from (\ref{AsGz}). 
The relations (\ref{Afinal})-(\ref{crossingt})
provide a system of coupled equations, which express
each amplitude $A_I(s,t) \,(I=0, 2)$ in terms of $\pi\pi$ and $\pi K$
$S$-wave phase shifts and the real
coefficients $c_n^{(I)}$  and $\tilde c_n^{(J)}\,(J=1/2\,,\, 3/2)$.

At fixed $t=m_\pi^2$ the amplitude (\ref{Afinal}) takes the simple form
\begin{equation}
\label{Atmpi2}
A_I(s, m_\pi^2)= \Omega_{I}(s) \left[\sum\limits_J C_{IJ} \tilde c_0^{(J)}\, 
\{ f_I(s)+\Omega_{I}^{-1}(s) \} +\sum\limits_n c_n^{(I)} [z(s)]^n\right] \,,
\end{equation}
where we introduced the last term of Eq. (\ref{Afinal}), {\em i.e.} $A_t^I(t)$ 
evaluated for $t=m_\pi^2$, inside the brackets, and expressed 
$A_t^I(m_\pi^2)$ according to Eq. (\ref{Atmpi1}).

It is easy to verify that the function multiplying the Omn\`es factor in
the relation (\ref{Atmpi2}) is real for $s<s_{in}$. Indeed,
the coefficients $c_n^{(I)}$ and $\tilde c_0^{(J)}$ are real, and for
values of $s$ below the inelastic threshold the variable  $z(s)$
is real. The only terms having an imaginary part for $s<s_{in}$ are the 
function $f_I(s)$ and the Omn\`es function $\Omega_{I}(s)$. 
But it is easy to check, using (\ref{Omegas})
and (\ref{f}), that their imaginary parts compensate in Eq. (\ref{Atmpi2}):
\begin{equation}
\label{compens}
{\rm Im}\,[ f_I(s)+\Omega_{I}^{-1}(s)]={\sin\delta_0^I(s)\over
|\Omega_I(s)|}-{\sin\delta_0^I(s)\over |\Omega_I(s)|}=0\,.\end{equation}
Therefore, the first term in the r.h.s. of (\ref{Atmpi2}) is real along the 
elastic region and can be included in the expansion in powers of the variable 
$z(s)$.
This shows that the general representation (\ref{Afinal})
reduces, when $t=m_\pi^2$, to  the Omn\`es representation (\ref{AGz}) derived
in the previous subsection. For arbitrary values of $t$, however, the amplitude
(\ref{Afinal}) has additional cuts in the elastic region of 
both the $s$ and $t$ channels.

The physical amplitude is obtained from (\ref{Afinal}) for $s=m_K^2$ and 
$t=m_\pi^2$. With our normalization, it depends on the $S$-wave phase shift
of $\pi\pi$ scattering and the Taylor coefficients $c_n^{(I)}$ and 
$\tilde c_n^{(J)}$ (the $S$-wave  phase shift of the 
$\pi K$ scattering contribute only indirectly, through these coefficients).
As proved in \cite{CiFi}, by the conformal mapping the rate of convergence 
is improved, so we expect to obtain an accurate representation at
low energies with a small number of terms in the expansions. In the next 
section we shall discuss how to determine the coefficients 
$c_n^{(I)}$ and $\tilde c_n^{(J)}$ using lattice results at unphysical points.

We end this section with two remarks: first we notice that in the above 
derivation  the symmetry between the two final pions is not explicit. 
This symmetry can be easily imposed by writing down a dispersion relation 
symmetrical with respect to the interchange of $t$ and $u$. Namely, instead 
of Eq. (\ref{Ast}) we have, more generally
\begin{equation}
\label{Astu}
A_I(s,t, u)=A_s^I(s)+{1\over 2} \left[A_t^I(t)+ A_u^I(u)\right]\,,
\end{equation}
where $A_u^I(u)$ satisfies a dispersion relation similar to (\ref{AtGw}).

The second  remark concerns the starting point used for the analytic 
extrapolation: in our analysis we considered the amplitude given by the 
expression (\ref{AI1}), which was obtained by making the LSZ reduction of the 
kaon in the $S$-matrix element (\ref{defS}). Alternatively, by reducing first
one pion  instead of the kaon, we obtain the  expression
\begin{equation}
\label{AI2}
A_I = {1\over \sqrt{2 k_{10}}}\langle  \pi(k_2) |
\,\eta_\pi(0)\,| K(p)\rangle\,.
\end{equation}
By further reducing the $K$-meson, one obtains, instead of (\ref{lsza}), 
the expression
\begin{equation}
\label{lsza1}
A_I(s,t) ={i\over \sqrt{4
k_{10} p_0}}\int {\rm d}x e^{-ip x} \theta (x_0) \,\langle
\pi(k_2)|[\eta_{\pi}(0), \eta_{K} (x)]|0\rangle\,,
\end{equation}
which allows the analytic continuation with respect to the variables  
$s=p^2$ and  $t=(p-k_2)^2$. It is easy to see that the spectral function 
can be written as in Eq.  (\ref{sigma}), with the corresponding terms  
similar to
(\ref{sigmas})-(\ref{sigmat}), excepted that the momentum conservation reads 
now $p_n=p$ in $\sigma_s$ and $p_n=k_2-p$ in $\sigma_t$. This implies that 
the significance of the various terms in the spectral functions  are 
different in the two approaches:
in Eq. (\ref{sigmas}) the c.m. energy which generates the intermediate states 
is yielded  by the total momentum carried by the kaon and
the interaction hamiltonian, while in 
(\ref{sigmat}) this energy is provided by the unphysical mass of the pion. 
In the alternative approach mentioned above, the energy in the term $\sigma_s$
is yielded by the unphysical mass of the kaon, while in
$\sigma_t$ it is provided by the momenta of the pion and of the hamiltonian.  
Despite this different interpretation of the matrix elements, it is easy to 
see that the Omn\`es representation is formally  similar in the two 
approaches. The differences concern only the inelastic contributions, 
parametrized by the expansions in powers of the conformal mapping variables.

\section{Discussion}

The Omn\`es representations derived above generalize previous results obtained
in the literature.
In Refs. \cite{PaPi,PaPi1} the authors write down an Omn\`es 
representation for the decay amplitude as a function of $s$ at fixed 
$t=m_\pi^2$, using the formal analogy with the scalar
pion form factor. The decay amplitude is identified, up to a constant, to
the Omn\`es factor $\Omega_{I}(s)$ in  Eq. (\ref{AGz}), {\em i.e.} the 
inelastic singularities are neglected. The role of the inelastic 
FSI and ISI contributions in the dispersion relation  was discussed in 
Ref. \cite{Suzu}, assuming that  the decay amplitude defined in Eq. (\ref{A})
satisfies a Mandelstam representation.

A similar line is followed in Ref. \cite{BuCo1}, where the weak Hamiltonian
${\cal H}_w$ in Eq. (\ref{A}) is
identified with a  field operator with the quantum numbers of the kaon,
and is assumed to carry a nonzero momentum. In this approach the
$K\to\pi\pi$ decay amplitude
is obtained from a dispersion relation for the elastic processes
$\bar KK\to\pi\pi$ ($s$-channel) and  $\pi K\to\pi K$ ($t$-channel). 
Assuming that
only $S$ and $P$  waves contribute to both channels, the amplitude is written 
in \cite{BuCo1} as a sum of functions depending on a single variable, 
$s$ ($t$), respectively. The interplay
between the weak and the strong dynamics is however not apparent in this 
treatment: the weak $K\to\pi\pi$ decay amplitude is given formally
by the amplitude of the strong scattering process $\bar KK\to\pi\pi$,
evaluated at an unphysical point.

As we mentioned already, the dispersion relations for the $K\to\pi\pi$ decay 
require a continuation in the external momenta. 
Therefore, the interpretation of the dispersive variables was not
very clear in the previous works. In ref. \cite{PaPi} the variable $s$ 
in the dispersion relation for the amplitude $A_I(s, t)$ at fixed 
$t=m_\pi^2$ was identified with the momentum squared of an off-shell kaon. 
This raised the subsequent criticism \cite{BuCo}, which
emphasized the ambiguity of the ChPT calculations for off-shell operators. 
We mention also that in Ref. \cite{Buras} the same variable $s$ was 
identified with the mass squared of the kaon.

In the present treatment, the LSZ reduction formula allows the analytic
continuation of the amplitudes in the complex planes of the external momenta.
No off-shell extrapolation of the operators is necessary and
the meaning of the variables is clear, as discussed at the end of the 
previous section.  If we use as starting point of the analytic continuation
the matrix element (\ref{AI1}), the variable
$s$ is defined  as $s=(k_1+k_2)^2$, and  in unphysical configurations it
may be different from the kaon momentum. 
Moreover, in this case the variable $t$ is equal to $k_1^2$, and represents 
the mass squared of the external pion.
On the other hand, if we use as  starting point of the analytic continuation 
the expression (\ref{AI2}), we have $s=p^2$, {\em i.e.} it represents the mass
squared of the external kaon, while $t=(p-k_2)^2$ includes,  in unphysical 
configurations, the momentum carried by the interaction hamiltonians.  
This interpretation allows us to incorporate in the dispersion relations, 
at least to a certain extent, the results of the lattice simulations. 
We consider for illustration the first definition of dispersive variables.

Most lattice calculations simulate matrix elements of the form
$\langle \pi|O_j|K\rangle$, related to the matrix elements of interest
$\langle \pi\pi|O_j|K\rangle$ by lowest order ChPT in the soft pion
limit \cite{Bern}. In the limit $k_1\to 0$ we have $t=k_1^2=0$
and $s=(k_1+k_2)^2=m_\pi^2$. Therefore, the lattice
simulations of this type provide the value of the amplitude $A_I(m_\pi^2, 0)$.

Direct simulations of the matrix elements $\langle \pi\pi|O_j|K\rangle$ are
done only for special configurations,  for instance when the kaon and one pion
are at rest. Assuming that the reduced pion is at rest, ${\bf k_1}=0$, we have
$s=t+m_\pi^2 + 2 \sqrt{t}\, E_\pi$, where $E_\pi$ is the energy of the moving 
pion.
At present the lattice simulations are done using values of the mass 
$\tilde m_\pi$ larger than the physical one.
To match this situation we take $\tilde t=  \tilde m_\pi^2 $, which means that
$\tilde s= \tilde m_\pi^2 + m_\pi^2+2 \tilde m_\pi\, E_\pi$. We assume 
therefore that $A_I(\tilde s, \tilde t)$ is known approximately from the 
lattice calculations\footnote{In the alternative interpretation of the 
dispersive variables, discussed above, the corresponding points
are $\tilde s=\tilde m_K^2$ and
$\tilde t=\tilde m_K^2 +m_\pi^2- 2 \tilde m_K\,E_\pi$.}.
Actually, $A_I(\tilde s, \tilde t)$  does not correspond
exactly to the configuration which is simulated on the lattice, since in the
dispersion relation the unreduced pion has the physical mass, $k_2^2=m_\pi^2$.
The analytic continuation with respect to $k_2^2$ requires the reduction of 
the second pion in the relation (\ref{lsza}), and is more complicated. In the 
present formalism, we can assume that the lattice situation is approached by 
using a dispersion relation  symmetrized with respect to the interchange of $t$
and $u$, as in Eq. (\ref{Astu}), where one pion has the physical
mass and  the other a different mass, $\tilde m_\pi$.
This is still not identical with the lattice
case, but we believe that even approximate numbers are welcome, since the 
dispersive approach includes correctly the FSI.  
Using a sufficient number of points $(\tilde s, \tilde t)$, it is possible to
determine  the coefficients $c_n^{(I)}$ and $\tilde c_n^{(J)}$ 
of the Taylor expansions in powers of the conformal variables $z$ and $w$, 
and to calculate then the physical amplitude using the expression
(\ref{Afinal}) for $s=m_K^2$ and  $t=m_\pi^2$.

\section{Conclusions}

In the present paper we derived Omn\`es
representations for the $K\to\pi\pi$ amplitudes, which include elastic and
inelastic contributions in both the direct and the crossed channels.
We showed that the amplitude is decomposed as a sum of
two functions, one depending only on $s$ and the other depending only on $t$.
This decomposition follows naturally from the LSZ  formalism and hadronic
unitarity and does not require additional assumptions.
The elastic contributions are parametrized by Omn\`es factors according 
to Watson theorem, and the inelastic singularities are accounted 
for by the technique of conformal mappings.
The treatment based on LSZ formalism allows a clear interpretation of the
dispersion relations and the meaning of the dispersive variables.
The unknown coefficients $c_n^{(I)}$ and $\tilde c_n^{(J)}$
entering  the parametrization (\ref{Afinal}) of the amplitude can be 
determined, at least approximately,  using information  provided by lattice
calculations at unphysical momenta.
The numerical implementation of this program is a subject of a future work. 
We mention finally that the effects of isospin violation, which were discussed
recently in Ref. \cite{Ciri}, can be incorporated in the dispersive treatment
by a suitable modification of the unitarity relation.
\vskip 0.3cm \noindent {\bf Acknowledgments:}  This work  was realized in the
frame of the Cooperation Agreement between CNRS and the Romanian Academy.
One author (IC) thanks the Centre de Physique Theorique de Marseille for
hospitality. We are grateful to G. Colangelo and L. Lellouch for many 
illuminating discussions.
The financial support of the Romanian Ministry of Education and Research under
Contract Ceres  Nr. 62/2001 is aknowledged.




\begin{thebibliography}{99}

\bibitem{ChPT} W. A. Bardeen, A. J. Buras, J.-M. G\'erard, Phys. Lett. 
B {\bf 192}, 138 (1987); J. Kambor, J. Missimer, D. Wyler, Phys. Lett.
B {\bf 261}, 496 (1991); J. Biknemns and J. Prades, JHEP {\bf 9901}, 023 
(1999); T. Hambye et al., Nucl. Phys. B {\bf 564}, 391 (2000)

\bibitem{Sach} C. T. Sachrajda, 20-th International Symposium
on Lepton and Photon Interactions at High Energies, Roma (July 2001), 
hep-ph/0110304;
G. Martinelli, Preprint ROMA-1322/01, hep-ph/0110023

\bibitem{Bern} C. Bernard et al., Phys. Rev. D {\bf 32}, 2343 (1985)

\bibitem{LeLu} L. Lellouch, M. L\"uscher, Commun. Math.  Phys. {\bf 219}, 
31 (2001)

\bibitem{Bouc} Ph. Boucaud et al., (SPQ(CD)R Collaboration), Nucl. Phys. Proc. 
Suppl. {\bf 106}, 329 (2002)

\bibitem{Lin} C.-J. Lin, G. Martinelli, E. Pallante, C.T. Sachrajda and 
G. Viladoro, hep-lat/0209020.

\bibitem{Truong} T. N. Truong, Phys. Lett. B {\bf 207}, 495 (1988)

\bibitem{PaPi} E. Pallante and A. Pich, Phys. Rev. Lett. {\bf 84}, 2568 (2000);
Nucl. Phys. B {\bf 592}, 294 (2000)

\bibitem{PaPi1} E. Pallante, A. Pich and
 I. Scimemi,  Nucl. Phys. B {\bf 617}, 441 (2001)

\bibitem{Omnes}  R. Omn\`es, Nuovo Cimento {\bf 8}, 316 (1958)

\bibitem{Buras} A. J. Buras et al., Nucl. Phys. B {\bf 565}, 3 (2000)

\bibitem{Suzu} M. Suzuki, Int. J. Mod. Phys. A {\bf 16}, 4637 (2001)

\bibitem{BuCo} M. B\"ucher, G. Colangelo, J. Kambor, F. Orellana,
 Phys. Lett. B {\bf 521}, 29 (2001)

\bibitem{BuCo1} M. B\"ucher, G. Colangelo, J. Kambor, F. Orellana,
 Phys. Lett. B {\bf 521}, 22 (2001)

\bibitem{Cola} G. Colangelo, Nucl. Phys. Suppl. {\bf 106}, 53 (2002)

\bibitem{LSZ} H. Lehmann, K. Symanzik,  W. Zimmermann, Nuovo Cimento,
{\bf 1}, 205 (1956); {\bf 2}, 425  (1957)

\bibitem{CaMi} I. Caprini, L. Micu, C. Bourrely, Phys. Rev. D {\bf 60}, 074016
 (1999); D {\bf 62}, 034016 (2000)

\bibitem{CaMi1} I. Caprini, L. Micu,  C. Bourrely, Eur. Phys. J. C {\bf 21}, 
145 (2001)

\bibitem{Musk} N.I. Muskhelishvili, Singular Integral Equations,
(Noordhoff-Groningen, 1953)

\bibitem{PhTr} T. N. Pham and T. N. Truong, Phys. Rev. D {\bf 16}, 896 (1977)

\bibitem{Hweak} A. J. Buras,  Weak interactions, CP Violation and Rare Decays,
in "Probing the Standard Model of Particle Interactions", Proc. 1997 Les 
HouchesSummer School, eds. R. Gupta  et al. 
(Elsevier, Amsterdam, 1999), Vol. I, p. 281

\bibitem{Bart} G. Barton,  Introduction to Dispersion Techniques in Field 
Theory (Benjamin, New York, 1965)

\bibitem{KaWi} G. K\"allen and A. S. Wightman, Dan. Vid. Selsk. Mat-Fys. Skr.
{\bf  1}, (1958) No. 6 
\bibitem{Watson} K. M. Watson, Phys. Rev. {\bf 95}, 316 (1958)

\bibitem{CiFi} S. Ciulli, J. Fischer, Nucl. Phys. {\bf 24}, 465 (1961)

\bibitem{HeLa} M. F. Heyn, C.B. Lang, Zeit. Phys. C {\bf 7}, 169 (1981);
I. Caprini, Eur. Phys. J. C {\bf 13}, 431 (2000);
J. F. de Troc\'oniz, F. J. Yndurain, Phys. Rev. D {\bf 65}, 093001 (2002)

\bibitem{AnBu} B. Ananthanarayan, P. Buttiker, Eur. Phys. J. C {\bf 19},
517 (2001)

\bibitem{Ciri} V. Cirigliano, J. Donoghue, E. Golowich, Phys.Rev. D {\bf 61}, 
093002 (2000)

\end{thebibliography}
\end{document}